# Seeded intermodal four-wave mixing in a highly multimode fiber

ABDELKRIM BENDAHMANE,[1*] KATARZYNA KRUPA,[1,2] ALESSANDRO TONELLO,[3] DANIELE MODOTTO,[2] THIBAUT SYLVESTRE,[4] VINCENT COUDERC,[3] STEFAN WABNITZ,[1,2] AND GUY MILLOT[1**]

[1]Université de Bourgogne Franche-Comté, ICB, UMR CNRS 6303, 9 Avenue A. Savary, 21078 Dijon, France
[2]Università di Brescia, Dipartimento di Ingegneria dell'Informazione, via Branze 38, 25123 Brescia, Italy
[3]Université de Limoges, XLIM, UMR CNRS 7252, 123 Avenue A. Thomas, 87060 Limoges, France
[4]Université Bourgogne Franche-Comté, Institut FEMTO-ST, UMR CNRS 6174, 15B Avenue des Montboucons, 25030 Besançon, France

*Corresponding authors: *abdelkrim.bendahmane@u-bourgogne.fr ; **guy.millot@u-bourgogne.fr ;

We experimentally and theoretically investigate the process of seeded intermodal four-wave mixing in a graded-index multimode fiber, pumped in the normal dispersion regime. By using a fiber with a 100 μm core diameter, we generate a parametric sideband in the C-band (1530–1565 nm), hence allowing the use of an Erbium-based laser to seed the mixing process. To limit nonlinear coupling between the pump and the seed to low-order fiber modes, the waist diameter of the pump beam is properly adjusted. We observe that the superimposed seed stimulates the generation of new spectral sidebands. A detailed characterization of the spectral and spatial properties of these sidebands shows good agreement with theoretical predictions from the phase-matching conditions. Interestingly, we demonstrate that both the second and the fourth-order dispersions must be included in the phase matching conditions to get better agreement with experimental measurements. Furthermore, temporal measurements performed with a fast photodiode reveal the generation of multiple pulse structures.

## 1. INTRODUCTION

The study of spatiotemporal nonlinear phenomena in multimode fibers (MMFs) has emerged as an active field of research over the past few years [1, 2]. Although step-index multimode fibers also permit ultra-wideband spectral translation via four-wave mixing [3, 4], many intriguing nonlinear optical effects involve graded-index multimode fibers (GRIN-MMFs), thanks to their very particular properties. In these fibers, the equal spacing of the modal wave numbers, in combination with their overall small modal dispersion, provide an ideal environment for the build-up of nonlinear effects [5-9]. Among the recently reported phenomena, we may cite the observation of multimode solitons and their associated dispersive waves [1, 10-12], geometric parametric instability or GPI (due to periodic self-imaging of the multimode beam) [13], self-induced beam cleanup [13-15], wideband supercontinuum generation [16-18], and coupled cavity laser emission [19].

For the special case of GPI, the observed sidebands grow from noise, which requires up to tens of kW peak pump powers to initiate the process [13]. An obvious solution to overcome this issue, and mitigate pump power requirements consists in stimulating the nonlinear mixing mechanism by adding a signal laser centered at the wavelength of the generated sidebands. Such a strategy is commonly used in single-mode fiber optical parametric amplifiers [20]. However, it is difficult to seed parametric processes in MMF, since the large frequency detuning of the corresponding sidebands is generally located in spectral regions hardly accessible by standard lasers. In order to tune MMF parametric sidebands into regions that are compatible with standard telecommunication sources, a simple solution would be to use a fiber with a properly adjusted core diameter. Indeed, as theoretically demonstrated by Longhi [21], and recently experimentally confirmed [22], the frequency detuning of the parametric sidebands is inversely proportional to the square root of the diameter of the fiber core.

Another important issue to be considered, when seeding parametric processes in MMFs, is the multimode nature of the propagating beams. This fact generally makes more difficult the analysis of the coupling processes between the different modes of the pump and the seed. To simplify the analysis, one may choose to excite low-order spatial modes only, for example by considering a pump that mainly propagates in the fundamental mode. In this situation, we may more easily interpret the mode interaction process in terms of intermodal four-wave mixing (IMFWM) [8, 23, 24].

In this paper, we experimentally investigate seeded IMFWM in a GRIN-MMF pumped in the normal dispersion regime at 1064 nm. We demonstrate that, in addition to IMFWM sidebands, seeding leads to the unexpected generation of new parametric sidebands in the visible and near-infrared regions of the spectrum. We employed a GRIN fiber with a 100 μm core diameter: as we shall see, this fiber is suitable to generate IMFWM spectral sidebands in the C-band (1530 – 1565 nm), which is accessible to wavelength tunable laser diodes and Erbium-based amplifiers. This paper is organized as follows. In section 2, we describe the experimental set-up and the optical properties of the fiber used. Next, we present in section 3 the experimental results of the IMFWM that are obtained in a standard, unseeded situation, where sidebands grow up from noise. We analyze the different mixing mechanisms, by providing a theoretical model for their frequency positions and by showing experimentally their spectral and spatial features. Then in section 4 we discuss the results obtained in presence of an additional seed. Finally, we report in section 5 real-time measurements, unveiling the presence of unexpected features in the temporal envelope of the generated sidebands.

## 2. EXPERIMENTAL SETUP

To experimentally investigate IMFWM, we used a 1-m-long piece of GRIN-MMF. Figure 1(a) shows the refractive index profile of the fiber, measured at 632 nm. It exhibits a parabolic shape in the fiber core that can be well modeled [black line in Fig. 1(a)] by the analytic formula: $n^2(\rho) = n_0^2 [1 - 2\Delta (\rho/R)^2]$, with $n_0 = 1.49$, $R = 53$ μm, $\Delta = 0.024$. For the pump laser, we used a Q-switched microchip Nd:YAG laser delivering 400 ps pulses at 1064 nm, with a repetition rate of 1 kHz. The maximum peak power of the emitted pulses was 220 kW. The polarization state and power of the pulses were controlled by the combination of a polarizer and two half-wave plates. Light was then coupled into the fiber through a 10 × microscope objective. To seed IMFWM, an additional amplified CW laser, continuously tunable around 1550 nm, was superimposed to the input pump wave with the help of a dichroic mirror. At the fiber output, the optical spectrum was recorded by use of three Yokogawa optical spectrum analyzers (OSAs) covering the range from 350 to 2400 nm. A CMOS camera was also used to spatially characterize the intensity profiles of the output beam, while temporal measurements were provided by means of a fast photodiode (rise time < 25 ps) and a real-time high bandwidth oscilloscope (50 GHz). Thanks to our set-up, we performed spectral, spatial and temporal measurements for a set of pump power values, both in the presence and in the absence of an additional seed wave. Let us first present results obtained without seeding the IMFWM process.

## 3. UNSEEDED CONFIGURATION

### A. Experimental results

Figure 2 depicts the evolution of the output spectrum, as the input pump peak power grows from 23.6 kW up to 70.5 kW. Although the fiber can support hundreds of guided modes, the pump injection conditions were adjusted in order to mainly excite the $LP_{01}$ fundamental mode at 1064 nm (Fig. 3(a)). For a peak power of 23.6 kW (blue spectrum), we observed the generation of a pair of sidebands around the pump at 1016 and 1116 nm (detuned by ±13.2 THz from the pump) due to the combined action of stimulated Raman scattering and intermodal modulational instability [25, 26]. We also observed the generation of a pair of highly detuned sidebands at

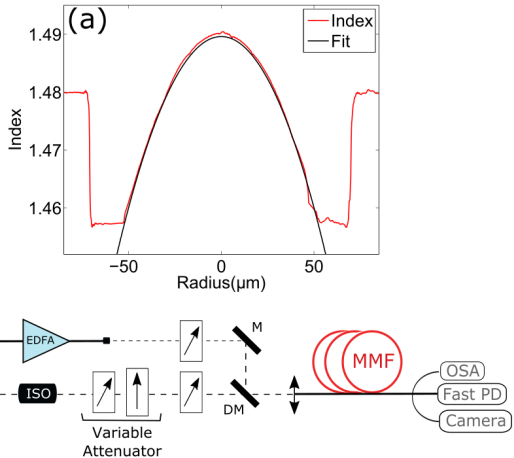

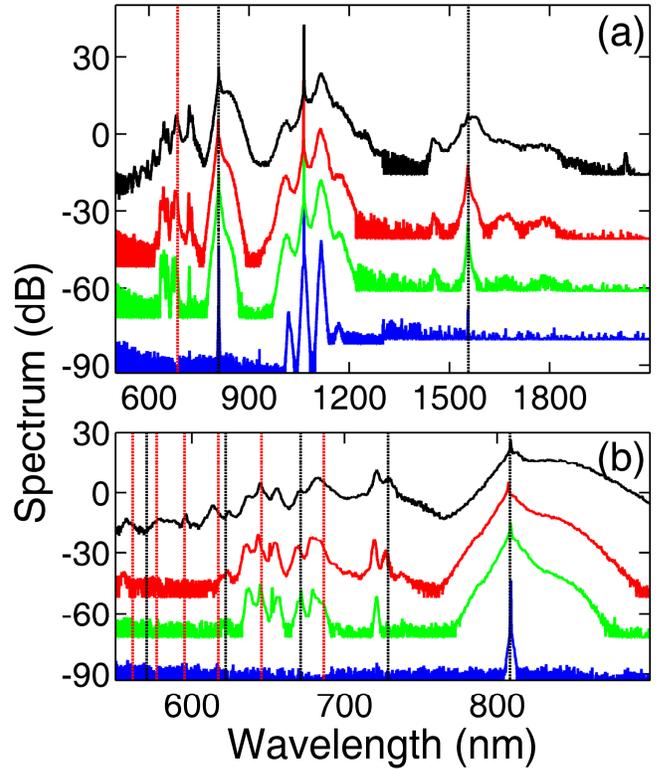

**Fig. 1.** (a) Refractive index profile measured at 632 nm. The black line indicates a parabolic fit of the core index profile (see text). (b) Experimental setup. TLS: tunable laser source, EDFA: Erbium doped-fiber amplifier; ISO: isolator; M: Mirror; DM: dichroic mirror; OSA: optical spectrum analyzer; PD: photodiode.

**Fig. 2.** (a) Output spectra recorded without seeding for a pump peak power of 23.6 kW (blue), 43.7 kW (green), 51.7 kW (red) and 70.5 kW (black). The vertical black (red) dashed lines indicate the position of the calculated first-order IMFWM sidebands generated by the primary (secondary) pump at 1064 nm (809 nm). (b) Zoom on the visible parts of the spectra. The vertical black (red) dashed lines indicate the position of the calculated IMFWM sidebands corresponding to the primary (secondary) pump (see text).

809 and 1555 nm (±89 THz). As discussed in section 3B, these sidebands can be ascribed to an IMFWM process, fed by the pump during its propagation. For increasing values of the peak power, we observed stronger conversion into the Raman and IMFWM sidebands. Also, additional weak Raman waves were generated by the 1555 nm component and small spectral peaks appeared between 630 and 730 nm (see green spectrum, Fig. 2). When we further raised the pump power above 51.7 kW (red spectrum, Fig. 2), the first-order IMFWM sideband at 809 nm begins to saturate, while its bandwidth progressively broadens, owing to Raman scattering. By increasing the pump power above 70 kW (black spectrum, Fig. 2), a stronger Raman broadening appears around the pump, while the conversion into the visible spectral sidebands is also enhanced, and many sidebands increase their spectral width. In the near-infrared part of the spectrum, we observed a broadening of the sideband at 1555 nm, and the appearance of a small sideband at 2027 nm.

## B. Sideband analysis

The frequencies of the parametric waves due to IMFWM can be obtained from the conservation laws of energy and propagation constant (phase matching condition) [5, 6, 8, 9]. Thus, for the special case of a degenerate pump (the two pump photons come from the same spatial mode) propagating in parabolic GRIN fiber, the frequency shift $f_n$ of the generated $n^{th}$-order IMFWM sideband can be predicted by the following analytical formula, which extends the result of Ref. [8] by including dispersion terms up to fourth-order:

$$f_n^2 \approx \frac{1}{4\pi^2}\left(\frac{-6\beta_2}{\beta_4} - \sqrt{\left(\frac{6\beta_2}{\beta_4}\right)^2 + \frac{24n\sqrt{2\Delta}}{R\beta_4}}\right) \quad (1)$$

where $\beta_2$ and $\beta_4$ are the second-order and fourth-order dispersion coefficients, respectively. At the pump wavelength $\lambda_p$=1064 nm the dispersion coefficients take the values $\beta_2 = 2.772\times10^{-26}\,s^2m^{-1}$ and $\beta_4 = -5.167\times10^{-56}\,s^4m^{-1}$. When applying Eq. (1) to the fiber under study, one obtains the wavelengths 808.5 nm and 1556.7 nm (±89 THz) for the first-order IMFWM sidebands. These wavelengths are indicated by vertical black dashed lines in Fig. 2(a). Their values are in very good agreement with our measurements (809 nm and 1555 nm). Note that neglecting fourth-order dispersion the frequency shift is given by $f_n^2 \approx \frac{1}{2\pi^2}\frac{n\sqrt{2\Delta}}{R\beta_2}$, leading to the wavelengths 813 nm and 1539 nm for the first-order sidebands. The above analysis shows that the fourth-order dispersion must be taken into account in the phase matching condition to accurately predict the wavelengths of the IMFWM sidebands. In the same way, we identified the small sideband observed at 2027 nm (visible in the black spectrum of Fig. 2(a)) as the second-order IMFWM Stokes sideband. Note also that, for the pump power levels used in our experiments (< 70.5 kW), the nonlinear terms in the phase-matching conditions are negligible and that therefore the sideband wavelengths do not depend on the pump power.

Figures 3 (a-c) display the recorded output transverse profiles of the spatial modes at wavelengths 1064, 809 and 1555 nm. Since we tuned the input conditions in order to obtain a pump mainly belonging to the $LP_{01}$ mode (Fig. 3(a)), selection rules impose that the generated sidebands should be composed of $LP_{0j}$ modes, where j varies from 1 to k+1 (k is an integer giving the order of the corresponding sideband) [23]. In the particular case of first order sidebands (k = 1), the Stokes line should either be generated in the $LP_{01}$ or in the $LP_{02}$ mode, while the anti-Stokes line should be guided in the $LP_{02}$ or $LP_{01}$ mode, respectively. The calculation of the nonlinear coupling among the involved spatial modes [8, 23] shows, that the most efficient frequency conversion should be that involving an anti-Stokes in the $LP_{01}$ mode and a Stokes in the $LP_{02}$ mode. Indeed as expected, the spatial mode recorded for the anti-Stokes sideband corresponds to $LP_{01}$ (Fig. 3(b)). However, the Stokes sideband does not perfectly match the expected profile of a $LP_{02}$ mode (Fig. 3(c)). This may originate from the presence of nearly degenerate guided modes that can spoil the original beam shape, owing to the presence of a fiber ellipticity or anisotropy. For instance the beam output observed in Fig. 3(c) can be obtained by combining the expected $LP_{02}$ with a fraction of power carried by the $LP_{21}$, which has a similar propagation constant.

Some of the sidebands belonging to the series of peaks observed in the visible region of the spectra are generated by higher-order IMFWM processes induced by the pump at 1064 nm. However, in a recent experimental work, involving a few-mode GRIN fiber with 22 μm core diameter [23], it has been demonstrated that, if the efficiency of energy transfer from the pump into the first-order anti-Stokes sideband is sufficiently high, a cascaded frequency conversion mechanism can take place. In fact, the anti-Stokes sideband plays the role of a secondary pump, and it generates new spectral sidebands by feeding its own IMFWM processes. From Eq. (1) with $\beta_2 = 4.799\times10^{-26}\,s^2m^{-1}$ and $\beta_4 = -1.162\times10^{-56}\,s^4m^{-1}$ at a wavelength of 809 nm, we may deduce that for the fiber under study, the first-order anti-Stokes sideband corresponding to this secondary pump should be located around 686 nm (detuned by +66.2 THz from 809 nm). This wavelength is indicated by a vertical red line in Fig. 2(a), which is close to the observed sideband at 680.6 nm (+70 THz from 809 nm). Note that the first-order Stokes wave induced by the secondary pump at an expected wavelength of 985 nm is not visible on the spectrum probably due to a too low intensity and to its superposition with the broad spectral components surrounding the main pump.

To further identify the spectral components associated with the primary (or main) pump and with the secondary pump, respectively, we represented in Fig. 2(b) the wavelengths corresponding to frequency shifts $f_n$ given by Eq. (1). The vertical black lines represented in Fig. 2(b) indicate the IMFWM sidebands associated with the primary pump, while the red lines correspond to the sidebands induced by the secondary pump. We notice that many of the observed components match well with the analytical prediction of Eq. (1), providing that we consider the corresponding pump.

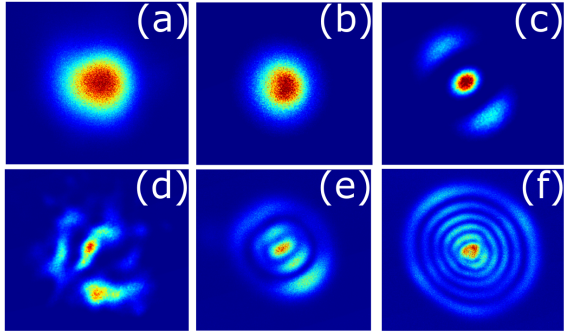

**Fig. 3.** Output beam profiles recorded at different wavelengths without (top level) and with seeding (lower level). (a) 1064 nm, (b) 809 nm, (c,d) 1555 nm, (e) 615 nm and (f) 554 nm using a 10 nm band pass filter.

## 4. SEEDED CONFIGURATION

### A. Experimental results

For the present GRIN-MM fiber, the first-order Stokes sideband at 1555 nm falls exactly in the C-band. Therefore the seeding process can be implemented with standard telecom sources. Thus in a second set of experiments we studied the configuration of seeded IMFWM, by superimposing to the primary pump an additional amplified CW-laser, centered close to the wavelength of the first-order Stokes, which is generated by IMFWM (at 1555 nm). The launching conditions of the seeding (or probe) light into the fiber were chosen by adjusting the transverse dimension, position and orientation of the beam at the input face of the fiber. Indeed, during our experiments, we observed that the largest spectra (or largest number of spectral sidebands) were obtained when the input beam profile of the seed was matching the entire fiber core, rather than just overlapping the pump profile (which was instead nearly single mode). Figure 3(d) displays the beam profile of the used seed at the MMF fiber output. We explain this fact by showing that the spatial profile of the spontaneously generated mode at 1555 nm (Fig. 3(c)) strongly differs from that of the pump. Note that at the wavelength of 1555 nm the camera detects mainly the CW seed (Fig. 3(d)), because its average power is much higher than that of the pulsed contribution.

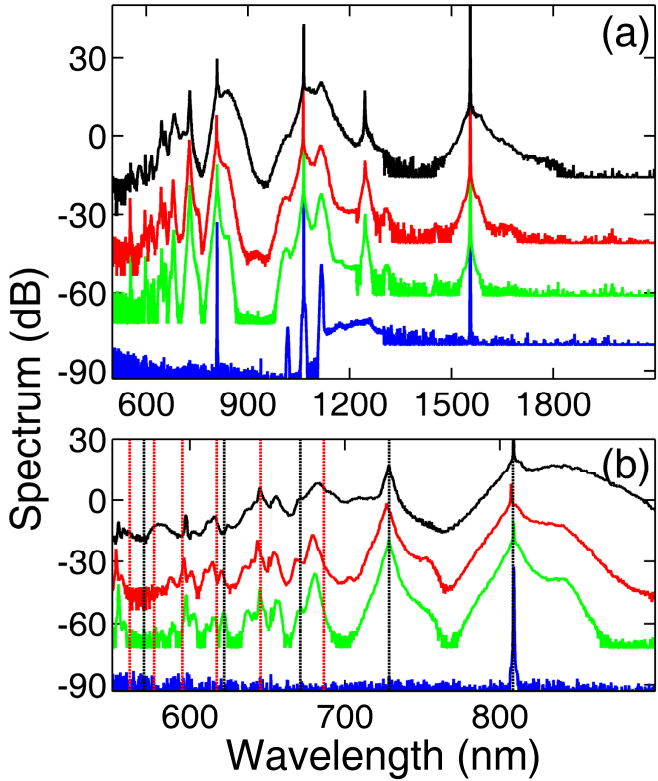

**Fig. 4.** (a) Output spectra recorded while seeding at 1555 nm for a pump peak power of 23.6 kW (blue), 43.7 kW (green), 51.7 kW (red) and 70.5 kW (black). (b) Zoom on the visible parts of the spectra. The vertical black (red) dashed lines indicate the position of the IMFWM sidebands corresponding to the primary (secondary) pump (see text).

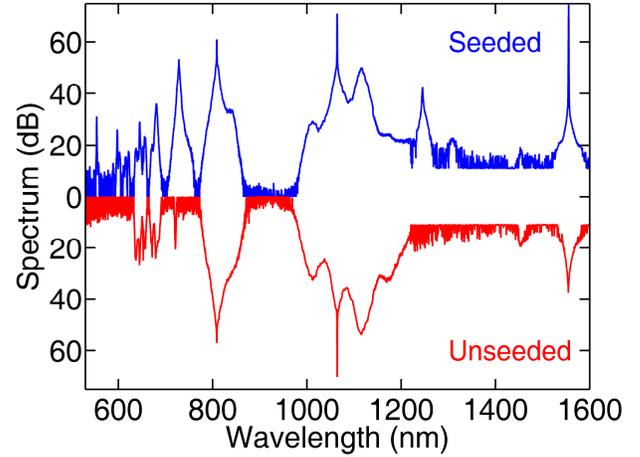

**Fig. 5.** Comparison of the output spectra recorded with seeding (blue) and without seeding (mirrored representation, red) for a pump peak power of 43.7 kW.

Figure (4) presents the evolution of the output spectrum, as the input pump peak power is increased from 23.6 kW up to 70.5 kW. The power of the CW laser was set to 100 mW, which corresponds to a fraction ranging from $1.4 \times 10^{-6}$ to $5 \times 10^{-6}$ with respect to the pump peak powers used in our experiments (see subsection B on the influence of the seed power). To improve the visual comparison, the spectra obtained with seed and without seed for a pump peak power P= 43.7 kW, are shown in Fig. 5. When comparing the seeded spectra with the unseeded spectra, we may easily notice the appearance of additional sidebands in the visible region, as well as in the near-infrared region at 1246 nm (-40.9 THz from 1064 nm). We also observe an improvement in the conversion efficiency into the sideband at 809 nm (on average, the increase is higher than 7 dB in peak intensity). However, the influence of the seed depends on the level of the pump peak power. As highlighted in Fig. 6, the seed influence is much more pronounced at moderate pump powers (< 50 kW, Fig. 6(a)), where many new sidebands are generated by the seeding, but it is less relevant at higher powers (70.5 kW, Fig. 6(b)). This reduced influence of the seed at high pump powers is due to the already strong level of pump depletion, and the associated intensity saturation of the sidebands in the visible region.

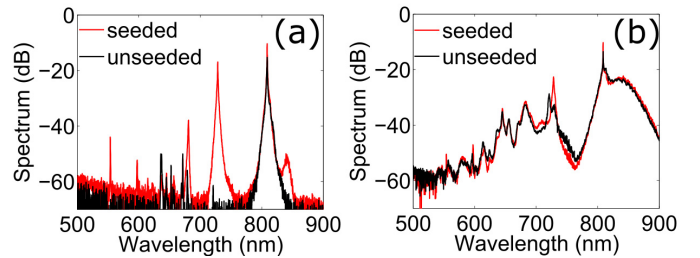

**Fig. 6.** Spectra recorded at two different pump peak powers: (a) 34.5 kW, (b) 70.5 kW in the seeded (red lines) and unseeded case (black lines). The seed power is set to 100 mW.

### B. Influence of the seed power

In addition to the previous experimental results, we also studied the influence of seed power on the dynamics of the frequency conversion processes. Figure 7 presents a selection of recorded spectra for different values of the seed power. All spectra were obtained by keeping the pump power fixed at 34.5 kW. Such value is within the range of pump power levels where the seed produces the largest influence. We note that a seed power of 5mW (green curve) generates a spectrum that is much richer in sidebands than the corresponding unseeded case (blue curve). At this seed power level we measured a parametric gain of 8 dB for the idler wave at 809 nm. However, by comparing the spectra represented by red and black curves, we may easily remark that, when the seed power exceeds a certain power level (*i.e.*, 15 mW), the influence of the additional increase of the seed power becomes largely irrelevant.

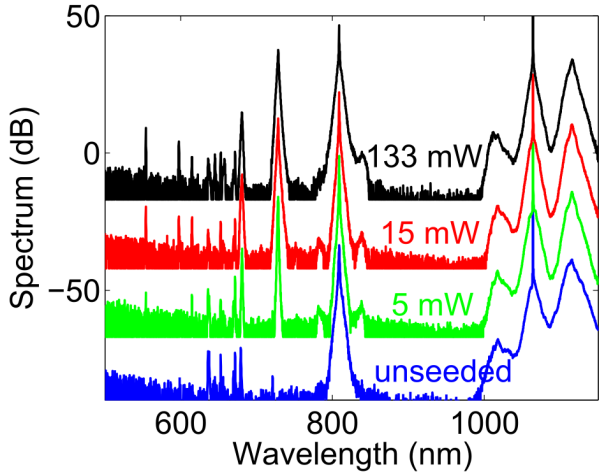

**Fig. 7.** Spectra recorded while seeding at 1555 nm with different seed powers (from 0 to 133 mW). The pump peak power was set to 34.5 kW.

### C. Sideband analysis

Similar to the unseeded case of Section 3, in order to identify the spectral components due to the primary and the secondary pumps, we represented in Fig. 4(b) the wavelengths corresponding to frequency shifts calculated from Eq. (1). Here also, the vertical black lines in Fig. 4(b) indicate the IMFWM sidebands generated by the main pump, whereas the red lines indicate sidebands generated by the secondary pump. As it occurs in the unseeded case (Fig. 2(b)), we may notice that most of the newly generated components follow qualitatively well the analytical square root law, although some of them are slightly shifted from the expected positions. We also remark that seeding increases the conversion efficiency into the frequency components generated by both primary and secondary pumps. Thanks to Eq. (1), we may identify the spectral band at 1246 nm seen in Fig. 4(a) as the 4th order Stokes sideband associated with the secondary pump. Figs. 3(e) and 3(f) display the recorded spatial modes profiles at 615 nm and 554 nm, respectively. These correspond to $LP_{03}$ and $LP_{06}$ modes, which is in agreement with the selection rule detailed in section 2.B. In fact, the 615 nm component corresponds to the 3rd order anti-Stokes sideband of the IMFWM process induced by the secondary pump, whereas the 554 nm component corresponds to the 6th order anti-Stokes sideband.

### 5. TEMPORAL MEASUREMENTS

In this section, we present the results of a temporal analysis of the generated sidebands, and compare results without and with seeding. Figure 8 illustrates the temporal profiles of pulses emerging from the 809 nm sideband in the seeded and unseeded case. As expected, the duration of the sideband pulse is shorter (170 ps) than the main pump pulse duration (400 ps) by a factor larger than two (black line, Fig.

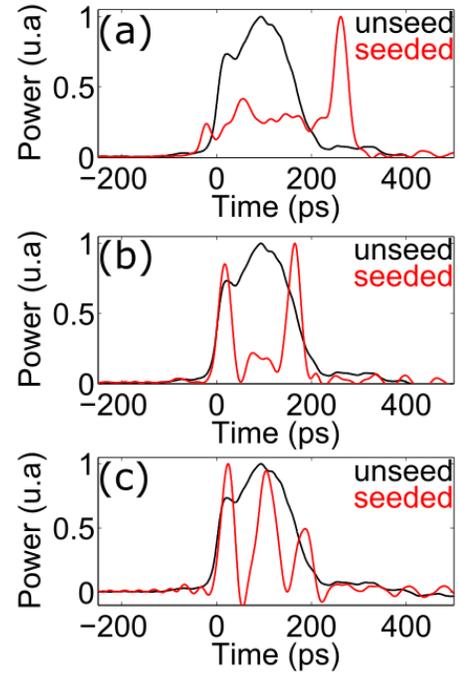

**Fig. 8.** Single-shot intensity-normalized temporal profiles measured in the unseeded (black lines) and seeded (red lines) configurations, respectively. The measurements have been done for a pump peak power set to 34.5 kW, while the frequency of the injected seed was tuned (see text).

8(a)). Moreover, seeding leads to a further time compression of the sideband pulses down to almost 35 ps, which matches with the time resolution of our detection system. Note that a large background is observed at the pulse leading tail. The resulting eleven-fold temporal compression results into a drastic increase in the pulse peak power, which increases the efficiency of spectral sidebands generation by the IMFMW process. More surprising is the fact that, when we slightly tuned the frequency of the seeding laser away from the IMFWM peak gain value (the frequency detuning was of the order of 10 GHz), we could observe a strong temporal reshaping of the sideband profile, leading to a deep temporal modulation of the sideband pulse or a multiple pulse structure, as shown in Fig. 8(b) and Fig. 8(c).

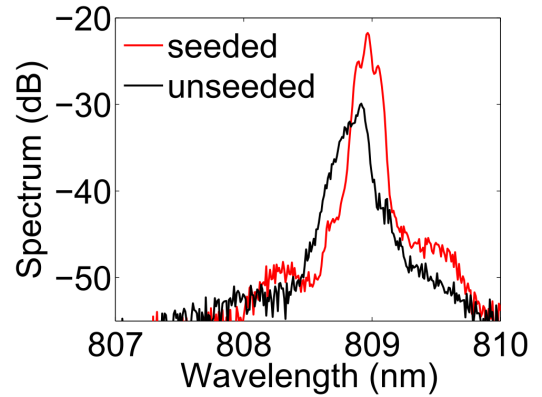

**Fig. 9.** Spectra recorded in the unseeded case (black line) and in presence of a detuned seeding laser (red line). The seed power is set to 100 mW.

Figure 9 depicts a close-up view of the 809 nm sideband spectrum in the unseeded (black line) and the seeded case (red line) associated to the triple pulse structure (Fig. 8(c)). A beating between the spontaneously generated sideband and the slightly detuned stimulated sideband may explain the intriguing temporal modulations observed. Let us note that these multi-pulse structures are sensitive to the input polarization and coupling conditions. On the other hand the intensities of the temporal modulations can vary from pulse to pulse. Further studies are required to understand their exact origin and behavior.

## 6. CONCLUSION

In summary, in this article, we presented an experimental study of the seeded IMFWM process in a highly multimode GRIN fiber. By reducing the waist diameter of the pump, we succeeded to limit the nonlinear interactions to the lowest-order fiber modes only. In these conditions, we studied the influence of a superimposed seed, centered on the first-order IMFWM Stokes sideband, on the efficiency of the multiple sideband generation processes. We showed that, for a certain range of parameters, the injected seed stimulates the generation of new spectral sidebands in the visible and near-infrared regions of the spectrum. We characterized the spectral positions and the spatial mode profiles corresponding to the newly generated sidebands, and found good agreement with theoretical predictions. We have demonstrated in particular that the fourth-order dispersion must be included in the phase matching conditions for better agreement with experimental measurements of the far-detuned parametric sidebands. Temporal measurements performed with the help of a fast photodiode and high-bandwidth real-time oscilloscope have revealed that seeding leads to a substantial increase in the achievable pulse compression factor. Frequency tuning of the seed away from the value leading to peak gain at low powers leads to the formation of multi-pulse structures, whose origin will be the subject of subsequent investigations.

These findings are of relevance for the development of new photonic devices such as parametric amplifiers and laser sources that expand the range of laser light emission via nonlinear frequency conversion processes in multimode silica or infrared optical fibers.

**Funding Information.** iXcore Research Foundation; Labex ACTION program (contract ANR-11-LABX-0001-01); Ministero dell'Istruzione, dell'Università e della Ricerca (MIUR) (PRIN 2015KEZNYM); Horiba Medical and BPI France within the Dat@diag project; The European Research Council (ERC) under the European Union's Horizon 2020 research and innovation programme (grant agreement No.740355). K.K. has received funding from the European Union's Horizon 2020 research and innovation programme under the Marie Sklodowska-Curie grant agreement No. GA-2015-713694 ("BECLEAN" project).